\documentclass[12pt,longbibliography]{revtex4-1}
\usepackage{ulem}
\usepackage{url}
\usepackage{epsfig}
\usepackage{graphicx,color}% Include figure files
\usepackage{epstopdf}
\usepackage[psamsfonts]{amssymb}
\usepackage{amsmath}
\usepackage{indentfirst}

\newcommand{\be}{\begin{equation}}
\newcommand{\ee}{\end{equation}}
\newcommand{\ba}{\begin{eqnarray}}
\newcommand{\ea}{\end{eqnarray}}

\begin{document}
\title{\large Given Enough Eyeballs, All Bugs Are Shallow? \\ Revisiting Eric Raymond with Bug Bounty Programs}

\author{Thomas Maillart}
\email{thomas.maillart@unige.ch}
\affiliation{University of Geneva, Geneva, Switzerland}

\author{Mingyi Zhao}
%\email{}
\affiliation{Pennsylvania State University, USA}

\author{Jens Grossklags}
\affiliation{Technical University of Munich, Germany}

\author{John Chuang}
\affiliation{University of California, Berkeley, USA}

\date{\today}

\begin{abstract}
\vspace{1cm}
Bug bounty programs offer a modern way for organizations to crowdsource their software security, and for security researchers to be fairly rewarded for the vulnerabilities they find. However, little is known on the incentives set by bug bounty programs -- how they drive engagement and new bug discoveries. This article provides an empirical investigation of the strategic interactions among the managers and participants of bug bounty programs, as well as the intermediation by bug bounty platforms. We find that for a given bug bounty program, each security researcher can only expect to discover a bounded number of bugs. This result offers a validation step to a theory brought forth early on by Brady et al. \cite{brady1999murphy}, which proposes that each security researcher inspecting a piece of software offers a unique environment of skills and mindset, which is amenable to the discovery of bugs that others may not be able to uncover. Bug bounty programs indeed benefit from the engagement of large crowds of researchers. Conversely, security researchers benefit greatly from searching for bugs in multiple bug bounty programs. However, we find that following a strong front-loading effect, newly launched programs attract researchers at the expense of older programs: the probability of finding bugs decays as $\sim 1/t^{0.4}$ after the launch of a program, even though bugs found later yield on average higher rewards. Our results lead us to formulate three recommendations for organizing bug bounty programs and platforms: (i) organize enrollment, mobility and renewal of security researchers across bounty programs, (ii) highlight and organize programs for front-loading, and (iii) organize fluid market transactions to reduce uncertainty and thus reduce incentives for security researchers to sell on the black market.

\end{abstract}

\maketitle

\section{Introduction}
\label{sec:intro}
On March 2nd, 2016, the Pentagon announced the launch of its first {\it bug bounty} program \cite{Pentagon}. From this point on, one of the most paranoid organizations in the world offered incentives to hackers to break into its systems and report found vulnerabilities for a reward. Although bug bounty programs have mushroomed in the last few years, this audacious announcement by a prominent defense administration may set a precedent, if not a standard, for the future of cybersecurity practice.\\ 

Software security has long been recognized as a challenging computational problem \cite{adams1984textordfeminineoptimizing} that often requires human intelligence. However, given the complexity of modern computer systems, human intelligence at the individual level is no longer sufficient. Instead, organizations are turning to tap the wisdom of crowds \cite{surowiecki2005wisdom} to improve their security. Software security is not alone. Other disciplines have similarly turned to mobilizing people at scale to tackle their hard problems, such as sorting galaxies in astronomy \cite{smith2013introduction}, folding proteins in biology \cite{khatib2011algorithm}, recognizing words from low quality book scans \cite{von2003captcha}, and addressing outstanding mathematics problems \cite{gowers2009massively,cranshaw2011polymath}. These examples involve different aspects of human intelligence, ranging from pattern recognition (e.g., Captcha \cite{von2003captcha}) to higher abstraction levels (e.g., mathematical conjectures \cite{gowers2009massively,cranshaw2011polymath}). It is not clear what kind of intelligence is necessary to find bugs and vulnerabilities in software, but it generally requires a high level of programming proficiency coupled with out-of-the-box thinking and {\it hacking skills} to find unintended uses for a software. \\

From hedonist pleasure to reputation building, to activism, motivations and incentives for hacking have evolved over time \cite{Levy84}. Among these, reputation and monetary incentives are increasingly put in place to entice security researchers to hunt for bugs. Bug bounty programs and online bug bounty platforms help set such incentives while facilitating communication and transactions between security researchers and software editors \cite{bohme2006comparison,finifter2013empirical,zhao2014exploratory,zhao2015empirical}. It however remains unclear how current mechanism designs and incentive structures will influence the long-term success of bounty programs. A better understanding of bug discovery mechanisms \cite{bishop1996conservative,brady1999murphy,zhao2016empirical}, and a better characterization of the utility functions of security researchers, organizations launching bug bounty programs and bug bounty platforms, will help shape the way bug bounty programs evolve in the future.\\

In this study, we investigate a public dataset of 35 public bug bounty programs from the {\it HackerOne} website \footnote{HackerOne : {\it http://www.hackerone.com}}. We find that with each vulnerability discovered within a bounty program, the probability of finding the next vulnerability decreases more rapidly than the corresponding increase in payoff. Therefore, security researchers rationally switch to newly launched bounty programs at the expense of existing ones. This switching phenomenon has already been reported in \cite{zhao2015empirical}. Here, we characterize it further by quantifying how incentives evolve as more vulnerabilities get discovered in a program and how researchers benefit in the long term by switching to newly launched programs. Our results help better understand the mechanisms associated with bug discovery, as they provide a validation step of the theory proposed in \cite{brady1999murphy}, and they help formulate concrete recommendations for the organization of both bug bounty programs and the online platforms supporting them.\\

This article is organized as follows. Related research is presented in Section \ref{sec:related}. Important features of the dataset are detailed in Section \ref{sec:data}. We introduce the main mechanism driving vulnerability discovery in Section \ref{sec:method}. Results are presented and discussed in Sections \ref{sec:results} and \ref{sec:discussion}, respectively. We offer concluding remarks in Section \ref{sec:conclusion}.

\section{Background}
\label{sec:related}

Software reliability is an age-old problem \cite{littlewood1973bayesian,adams1984textordfeminineoptimizing,littlewood1989predicting}. Early empirical work on software bug discovery dates back to the time of UNIX systems \cite{miller1990empirical}, and over the years, numerous models for vulnerability discovery have been developed (see \cite{avgerinos2014enhancing,zhao2016empirical} for some contemporary approaches). As early as in 1989, it was recognized that the time to achieve a given level of software reliability is inversely proportional to the desired failure frequency level \cite{adams1984textordfeminineoptimizing}. For example, in order to achieve a $10^{-9}$ probability of failure, a software routine should be tested $10^{9}$ times. Actually, the random variable $P(T > t) = 1/t$ corresponds to Zipf's law \cite{maillart2008empirical,saichev2009theory}, which diverges as the random variable sample increases (i.e., no statistical moment is defined). Thus, there will always be software vulnerabilities to be discovered as long as enough resources can be provided to find them.\\

Taking an evolutionary perspective brings additional insights. Finding bugs is comparable to the survival process involved in the selection of species: defects are like genes, which get expressed under the pressure of environment changes. Brady et al. \cite{brady1999murphy} have shown that software testing follows the principle of entropy maximization, which preserves {\it genetic variability} and thus, removes only the minimum possible number of bugs, following the exploration of use cases (i.e., the software environment). With their out-of-the-box -- {\it hacker mindset} -- thinking, security researchers are precisely good at envisioning a broad range of possible use cases, which may reveal a software defect (i.e., program crash) or an unintended behavior.\\

Software solutions have been developed to systematically detect software inconsistencies and thus potential bugs (e.g., Coverity, FindBugs, SLAM, Astree, to name a few). However, to date, no systematic algorithmic approach has been found to detect and remove bugs at a speed that would keep pace with software evolution and expansion. Thus, human intelligence is still considered as one of the most efficient ways to explore novel use case scenarios -- by manual code inspection or with the help of bug testing software -- in which software may not behave in the intended way.\\

Management techniques and governance approaches have been developed to help software developers and security researchers in their review tasks, starting with pair programming \cite{hulkko2005multiple}. To protect against cyber-criminals, it is also fashionable to hire {\it ethical hackers} -- who have a mindset similar to potential attackers -- to probe the security of computer systems \cite{smith2002ethical,saleem2006ethical,bishop2007penetration}. In this context, the policy of full disclosure, originating from the hacking and open source communities, plays a significant role in software security by forcing software owners to acknowledge and fix vulnerabilities discovered and published by independent researchers \cite{arora2008optimal}. The full-disclosure model has evolved into responsible disclosure, a standard practice where the security researcher agrees to allow a period of time for the vulnerability to be patched before publishing the details of the uncovered flaw. In most of these successful human-driven approaches, there is a knowledge-sharing component, either between two programmers sitting together in front of a screen, ethical hackers hired to probe the weaknesses of a computer system, or the broader community being exposed to open source code and publicly disclosed software vulnerabilities \cite{cavusoglu2007efficiency}. Thus, Eric Raymond's famous quote ``Given enough eyeballs, all bugs are shallow'' \cite{raymond1999cathedral} tends to hold, even though in practice things are often slightly more complicated \cite{hafiz2015game}.\\

One way to gather enough eyeballs is to recruit a larger crowd of security researchers. For this purpose, bug bounty programs and vulnerability {\it markets} have emerged in recent years to facilitate the trading of bugs and vulnerabilities. These two-sided markets provide economic incentives to support the transfer of knowledge from security researchers to software organizations \cite{camp2004pricing}, as they help simultaneously to harness the wisdom of crowds and to reveal the security level of organizations through a competitive incentive mechanism \cite{schechter2002buy}. Nonetheless, the efficiency of bug bounty programs has been questioned on both theoretical \cite{kannan2005market,mckinney2007vulnerability} and empirical grounds \cite{ransbotham2008markets,algarni2014software}.\\

Building on previous work by Schechter \cite{schechter2002buy}, Ozment \cite{ozment2004bug} theorized that the most efficient mechanisms are not markets {\it per se}, but rather auction systems \cite{milgrom1982theory}. In a nutshell, the proposed (monopsonistic) auction mechanism implies an initial reward $R(t=t_0) = R_0$, which increases linearly with time. If a bug is reported more than once, only the first reporter receives the reward. Therefore, security researchers have an incentive to submit a vulnerability early (before other researchers submit the same bug), but not too early, so that they can maximize their payoff $R(t) = R_0 + \epsilon \times t$ with $\epsilon$ the linear growth factor, which is meant to compensate for the increasing difficulty of finding each new bug. However, setting the right incentive structure $\{R_0,\epsilon \}$ is non-trivial given uncertainties in the amount of work needed, the level of competition (e.g., the number of researchers enrolled) in the bug bounty program \cite{pandey2014assessment}, or the nature and likelihood of overlap between two submissions by different researchers. Nevertheless, bug bounty programs have emerged as a tool used by many software organizations, with a range of heterogeneous incentive schemes \cite{finifter2013empirical}. For instance, some bug bounty programs include no monetary rewards \cite{zhao2014exploratory}. Meanwhile, dedicated platforms have been launched to act as trusted third parties in charge of clearing transactions between organizations and security researchers. These platforms also assist organizations in the design and deployment of their own program. One of the leading platforms is HackerOne, which runs public and private programs for organizations across a wide range of business sectors. A subset of the private and public programs award bounties, while other bug bounty programs capitalize on incentives associated with reputation building, which is an important motivation driver in the hacker community \cite{lakhani2005htu}. These programs report bounty awards on their company program pages on the HackerOne website. Previous research has investigated vulnerability trends, response and resolve behaviors, as well as reward structures of participating organizations \cite{zhao2014exploratory,zhao2015empirical}. In particular, it was found that a considerable number of organizations experienced diminishing trends for the number of reported vulnerabilities, even as the monetary incentives exhibit a significantly positive correlation with the number of vulnerabilities reported \cite{zhao2015empirical}. Recent research has also proposed approaches to improve the overall effectiveness of bug bounty programs and platforms by reducing the number of low-quality submissions, and by tackling the problem of duplicate submissions via novel incentive and allocation mechanisms, respectively \cite{laszka2016banishing, zhao2016crowdsourced, zhao2017devising}.

\section{Data}
\label{sec:data}
\begin{figure}[h]
\begin{center}
\includegraphics[width=12cm]{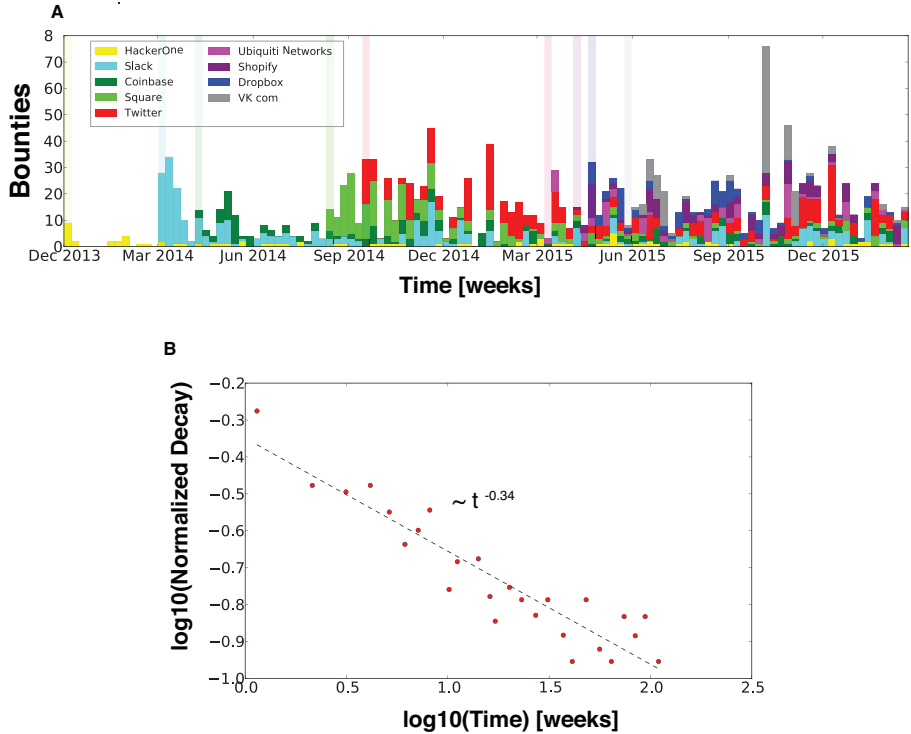}
\caption{\footnotesize {\bf A.} Weekly vulnerability discoveries for the 9 most active programs (with at least 90 bug discoveries as of February 15, 2016). The light colored vertical bars represent the start of the program, occurring when the first bounty is awarded. Most programs exhibit an initial shock, followed by a decay of discoveries, which is characterized at the aggregate level by a long-memory process (panel {\bf B}) characterized by a power law decay $\sim t^{\alpha}$ with $\alpha = -0.40(4)$ ($p < 0.001$ and $R^2 = 0.79$, obtained by ordinary least square fitting). Each data point in the figure is the median of normalized vulnerability numbers of all 35 programs considered in this study.}
\label{timeline}
\end{center}
\end{figure}

The data were collected from the public part of the HackerOne website. From 35 public bounty programs, we collected the rewards received by security researchers (in US dollars), with their timestamps (45 other public bounty programs do not disclose detailed information on rewards, and the number of private programs is not disclosed). Since HackerOne started its platform in December 2013, new public programs have been launched roughly every two months, following an essentially memoryless Poisson process ($\lambda = 57$ days, $p < 0.001$ and $R^2 > 0.99$). Figure \ref{timeline}A shows the timeline of the 9 most active programs with at least 90 valid (i.e., rewarded) bug discoveries, as of February 15th, 2016. When a new program is launched, we observe an initial peak within weeks after launch, which accounts for the majority of discoveries. After the initial surge of vulnerability discoveries, bounty awards become less frequent following a robust power law decay $\sim t^{\alpha}$ with $\alpha = -0.40(4)$ ($p < 0.001$ and $R^2 = 0.79$; the fit of the time series was obtained using ordinary least square regression as described in \cite{maillart2011quantification}) at the aggregate level and over all 35 bounty programs (see Figure \ref{timeline}B). Some programs depart from this averaged trend: For instance, Twitter exhibits a steady, almost constant, bug discovery rate and VKontakte exhibits its peak activity months after the initial launch. These peculiar behaviors may be attributed to program tuning and marketing, to sudden change of media exposure or even to fundamental differences of program comparative fitness, for which we do not have specific information.\\

The power law decay observed here is reminiscent of the long-established $\sim 1/\tau$ law of bug discovery in software testing \cite{adams1984textordfeminineoptimizing}. This similitude is interesting even though bug bounty programs do not provide direct information regarding software reliability. The difference of exponent ($\alpha \approx 0.4$ instead of $1$), may stem from long-memory processes associated with human behaviors and human timing effects, which correspond to task priority queueing and a rationale use of time as non-storable scarce resource \cite{maillart2011quantification}. Long-memory processes observed in collective human behaviors such as those observed on Figure \ref{timeline}B may also be associated with critical cascades of productive events \cite{sornette2014much}. The intuition is that each security researcher will generate a cascade of bug discoveries (of a size related to the total number of bugs discovered by this person, which is a random variable across all researchers), and by her activity (each researcher will influence and attract other researchers), hence generating cascades of joining and of bug discoveries.\\

%Here, the rationale for cascades builds on 2 ingredients : (i) each security researcher generates a cascade of bug discoveries, and (ii) each bug discovery attracts new security researchers, who in turn trigger their own bug discovery cascade. Cascades are typically slightly heavy-tailed, however bounded, as we'll later : we will dissect these cascades in conjunction with incentives provided by the accumulation of bounties (reputation, slight chance to find less obvious bugs bring more reputation, and increased bounties rewards)}.

Here, we do not consider the human timing effects encompassing delays, effort, processing time, and influence. We only consider incremental valid bug discovery and reporting by security researchers.

\section{Method}
\label{sec:method}
Bug bounty programs work on the premise that humans as a crowd are efficient at searching and finding bugs. Their mere existence is a {\it de facto} recognition that market approaches for bug discovery bring efficiency, beyond in-house security. Bug bounty programs signal that organizations are ready to complement their vertical cost-effective security operations with market approaches, which are traditionally perceived as less cost-effective, yet more adaptive \cite{coase1937}. Early on, Brady et al. \cite{brady1999murphy} have offered a hint for the existence of such markets for bugs: according to their proposed theory, each researcher has slightly different skills and mindset. When a security researcher tests a software piece by choosing the inputs, she offers a unique operational environment. This environment is prone to the discovery of new bugs, which may not have been seen by other researchers. The proposed theory by Brady et al. \cite{brady1999murphy} intrinsically justifies the existence of bug bounty program structures as markets, which provide the necessary diversity to account for the highly uncertain risk horizon of bug discovery. Here, we develop a quantitative method to formalize a mechanism and to test the theory proposed in \cite{brady1999murphy}. This validation step shall help provide organizational design insights for bug bounty programs.\\

For that, we investigate the interplay between the vulnerability discovery process and the cumulative rewards distributed to security researchers within and across 35 public bounty programs hosted on HackerOne. When a bug bounty program starts, it attracts a number of security researchers, who in turn submit bug reports. Subsequent bug discoveries get increasingly difficult for each individual researcher, and to some extent for all researchers together. The difficulties faced by security researchers can be technical. They can also be the result of insufficient or conflicting incentives. Here, we develop and test a model, which accounts for both technical difficulties and insufficient incentives. We further address conflicting incentives by measuring the effect of newly launched bug bounty programs on incumbent programs.\\

Starting from an initial probability of discovering the first vulnerability $P(k=0) = 1$, the probability to find a second bug is a fraction of the former probability: $P_{k+1} = \beta \cdot P_k$ with $\beta$ a constant strictly smaller than one. The probability that no more discoveries will be made after $k$ steps is given by $P_k = \beta^{k} (1-\beta)$. Conversely, starting from the initial reward $R_0 = R(k=0)$, the subsequent reward $R_1 = \Lambda_1 \cdot R_0$, and further additional reward $ R_2 = \Lambda_2 \Lambda_1 \cdot R_{0}$. After $n$ steps, the total reward is the sum of all past rewards: 

\begin{equation}
R_{n} = R_{0} \sum_{k=1}^{n} \Lambda_1 ... \Lambda_k.
\end{equation}

Thus, $R_{n}$ is the recurrence solution of the  
Kesten map ($R_{n} = \Lambda_n R_{n-1} +R_0$)
\cite{kesten1973random,sornette1997convergent}:  as soon as amplification occurs (technically, 
some of the factors $\Lambda_k$ are larger than $1$), the distribution
of rewards is a power law, whose exponent $\mu$ is a function of $\beta$
and of the distribution of the factors $\Lambda_k$. In the case where
all factors are equal to $\Lambda$, this model predicts three possible regimes for the distribution of rewards (for a given program): thinner than exponential for $\Lambda < 1$, exponential for $\Lambda = 1$, and power law for $\Lambda > 1$ with exponent $\mu = |\ln \beta|/ \ln \Lambda$ \cite{sornette2013exploring}. The expected payoff of vulnerability discovery is given by,

\begin{equation}
\label{ }
U_k = P_k \times R_k,
\end{equation}

\noindent with both $P_k$ and $R_k$ random variables respectively determined by $\beta$ and $\Lambda$. Because $U_k$ is a multiplication of diverging probability and reward components, its nature is reminiscent of the St. Petersburg paradox (or St. Petersburg lottery), proposed first by the Swiss Mathematician Nicolas Bernoulli in 1713, and later formalized by his brother Daniel in 1738 \cite{bernoulli1954exposition}. The St. Petersburg paradox states the problem of decision-making when both the probability and the reward are diverging for $k \rightarrow \infty$: a player has a chance to toss a fair coin at each stage of the game. The pot starts at two and is doubled every time a head appears. The first time a tail appears, the game ends and the player wins whatever is in the pot. Thus, the player wins two if a tail appears on the first toss, four if a head appears on the first toss and a tail on the second, eight if a head appears on the first two tosses and a tail on the third, and so on. The main interest of Bernoulli was to determine how much a player would be ready to pay for this game, and he found that very few people would like to play this game even though the expected utility increases (in the simplest case proposed by Bernoulli, $U_n = \sum_{k=0}^{n} U_k = n$) \cite{bernoulli1954exposition}. The situation of a security researcher differs from the St. Petersburg lottery as bug search costs are incurred at every step. Since these costs influence the probability to find an additional bug, they can be at least partially integrated in $P_k$. We could assume equivalently that costs are integrated into a net utility as $U^{*}_k = U_k - c_k$. Here, we do not factor these costs in because their exact nature is largely undetermined and our data do not offer a reliable proxy. The security researcher may also decide to stop searching for bugs in a program, at any time $k$: this is equivalent to setting $P_{k+1} = 0$.\\

The expected payoff $U_k$ therefore determines the incentive structure for security researchers, given that the bounty program manager can tune $R_0$ and to some extent $\Lambda$. The utility function may also incorporate non-monetary incentives, such as reputation: finding a long series of bugs may signal some fitness for a bug bounty program and thus create a permanent job opportunity \cite{moussouris2016}. Similarly, discovery of a rare (resp. critical) bug that no other researcher has found before has a strong signaling effect, which can help make a career. However, these strategies are high-risk high-return. Therefore, they result in additional fame. In the next section, we will calibrate our model to the bug discovery process associated with 35 bounty programs publicly documented on the HackerOne platform.

\section{Results}
\label{sec:results}
The discovery process in a bug bounty program is driven by the probability to find an additional bug given that $k$ bugs have already been discovered. Program managers aim to maximize the total number of bugs found $B_c$. Our results show that the number of bugs discovered is a super-linear function of security researchers who have enrolled in the program (see Figure \ref{fig:scaling}A). While bug bounty programs benefit from the work of a large number of researchers, researchers overall benefit from diversifying their efforts across programs (see Figure \ref{fig:scalings_awards}C). This benefit is particularly tangible regarding the cumulative reward they can extract from their bug hunting activity. In particular, we illustrate how researchers take the strategic decision to enroll in a newly launched program, at the expense of existing ones they have formerly been involved in.
 
\subsection{Security researcher enrollment determines the success of a bug bounty program}
\label{sec:enrollment}

\begin{figure}[ht]
\begin{center}
\includegraphics[width=12cm]{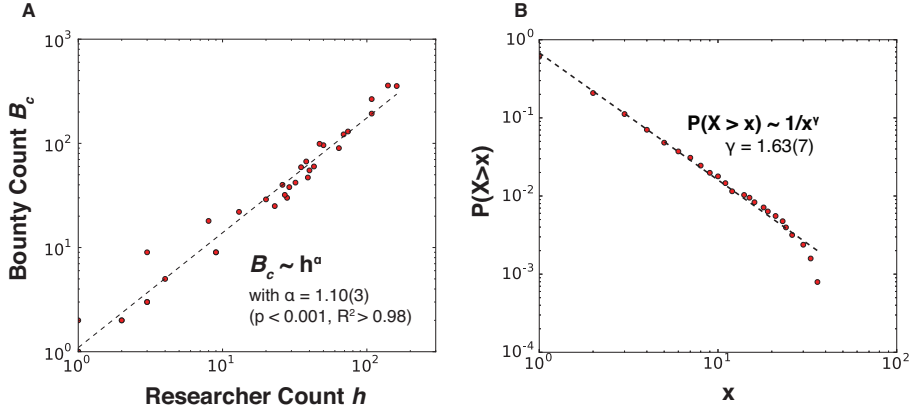}
\caption{\footnotesize {\bf A.}  The number of bounty discoveries per program $B_c$ scales as $~h^{\alpha}$ with $\alpha = 1.10(3)$ and $h$ the number of security researchers enrolled in a program (fit and confidence interval were obtained by ordinary least squares of the logarithm of researcher and bounty counts). Since $\alpha > 1$, a bounty program benefits in a super-linear fashion to the enrollment of more researchers. {\bf B.} The tail distribution of bounty discoveries per researcher per program follows a power law distribution $P(X>x) \sim 1/x^\gamma$  with $1 < \gamma = 1.63(7) < 2$ (obtained by maximum likelihood estimation and confidence interval bootstrapping, following \cite{maillart2008empirical,Clauset2009PowerLaw}). The distribution is therefore relatively well bounded (with the first moment being well-defined). Furthermore, we observe an upper cut-off of the tail with $x_{max} \approx 400$ bounties. Thus, from {\bf A.} and {\bf B.} combined, we find that the number of vulnerabilities is mainly driven by the number of researchers enrolled in programs.}
\label{fig:scaling}
\end{center}
\end{figure}

As captured in Figure \ref{fig:scaling}A, we find that the number $B_c$ of bugs discovered in a bug bounty program scales as $B_c \sim h^{\alpha}$ with $\alpha = 1.10(3)$ and $h$ the number of security researchers enrolled in a program. Since $\alpha > 1$, a bounty program benefits in a super-linear fashion from the enrollment of more researchers. This result is reminiscent of productive bursts and critical cascades of contributions in open source software development \cite{sornette2014much}: each enrollment (i.e., {\it mother} event) initiates a cascade of bug discoveries (i.e., {\it daughter} events). Here each cascade stems from a single security researcher and the nature of these cascades is captured at the aggregate level by their size as a random variable. As shown on Figure \ref{fig:scaling}B, the distribution of bounty discoveries per researcher and per program follows a power law tail $P(X>x) \sim 1/x^\gamma$  with $ \gamma = 1.63(7)$. The first moment of the distribution (i.e., the mean) is however well-defined (as a result of $ \gamma > 1$). Moreover, we observe an upper cut-off of the tail with $x_{max} \approx 40$ bounties. Thus, each enrollment of a security researcher in a program provides a statistically bounded amount of new bug discoveries.\\

\subsection{Security researchers are incentivized to diversify their contributions across bug bounty programs}
\label{sec:diversify}

\begin{figure}[ht]
\begin{center}
\includegraphics[width=17cm]{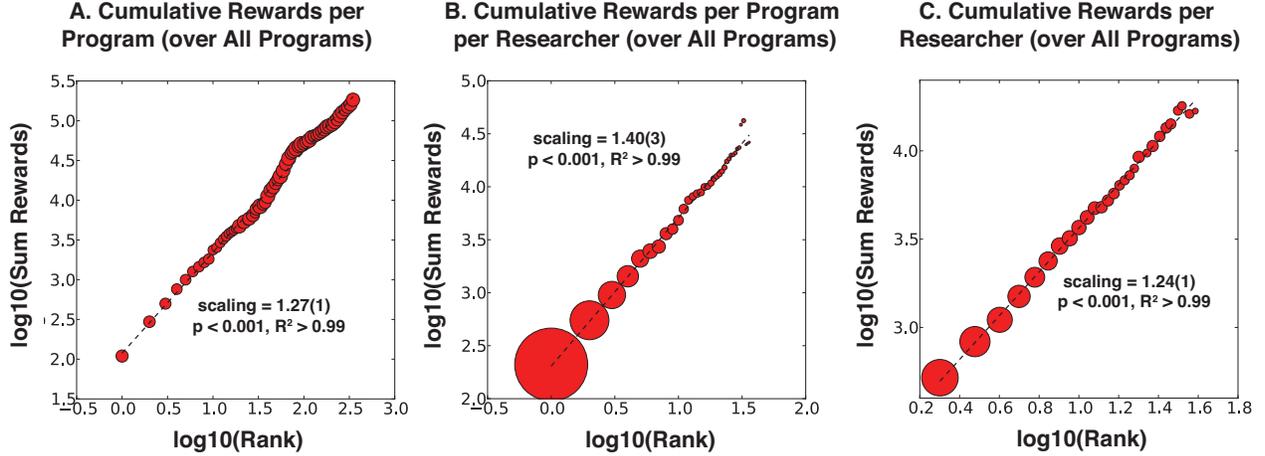}
\caption{\footnotesize {\bf A.} (Log-)binned cumulative down-payment per program over all public programs on the HackerOne platform, scales as $R_{k} \sim k^{1.27}$  with $k$ the rank ($ p < 0.001$, $R^2 > 0.99$; fit obtained with ordinary least squares of the logarithmic values on both axis). Each log-bin shows the mean value and the circle sizes depict the number of values in each bin (i.e., the rank frequency). The super-linear {\it scaling} relationship between the cumulative reward and the rank shows that reward increases as a function of $k$. However, the frequency of vulnerabilities $P_k$ is only slightly upwards trended increasing as $ \sim k^{0.13}$ ($ p < 0.001$, $R^2 = 0.40$). {\bf B.} Considering the security researcher's expected payoff for one bug bounty program, the super-linear effect is much stronger ($R_{k} \sim k^{1.40}$ with $ p < 0.001$ and $R^2 > 0.99$). However, the frequency decays following a power law of the form $P_{k} \sim k^{-1.85}$ ($ p < 0.001$, $R^2 = 0.97$). {\bf C.} Over all bug bounty programs, security researchers have another expected payoff: the reward scaling is smaller ($R_{k} \sim k^{1.24}$ with $ p < 0.001$, $R^2 > 0.99$), yet the frequency of bug discoveries decays much slower as a function of rank $P_{k} \sim k^{-1.05}$ ($ p < 0.001$, $R^2 = 0.85$).}
\label{fig:scalings_awards}
\end{center}
\end{figure}

For security researchers, the main metric is the expected cumulative payoff earned from the accumulation of bounty awards over all programs. This expected payoff is governed by the probability to find a given number of bugs and their associated payoffs, as discussed in Section \ref{sec:method}. To fully understand the incentive mechanisms at work, we consider three perspectives: (i) the expected cumulative down-payment made by bug bounty program managers (see Figure \ref{fig:scalings_awards}A), the expected cumulative payoff from the viewpoint of a security researcher for (ii) one program (see Figure \ref{fig:scalings_awards}B), and for (iii) all programs (see Figure \ref{fig:scalings_awards}C).\\

The average cumulative down-payment per program exhibits a super-linear scaling as $\sim k^{1.27}$ ($ p < 0.001$, $R^2 > 0.99$), while the frequency of bugs $P_k$ is only slightly upwards trended, increasing as $ \sim k^{0.13}$ ($ p < 0.001$, $R^2 = 0.40$). The expected down-payment by bug bounty program managers therefore scales as $\sim k^{1.40}$. This is a considerable super-linear increase (as $k \rightarrow \infty$), which casts questions on the long-term sustainability of bug bounty programs. \\

From the security researcher's viewpoint and her expected payoff from a single bug bounty program, the increase of average cumulative reward ($R_{k} \sim k^{1.40}$) does not offset the fast decay of probability ($P_{k} \sim k^{-1.85}$) to find a vulnerability of rank $k$. The expected payoff therefore follows $U_k \sim k^{-0.45}$, which does not bring high incentives to explore in depth a bug bounty program. However, it is important to note that the bug bounty manager cannot directly fix $P_k$, which can only be influenced -- to some extent -- by increasing rewards. To maintain positive individual incentives, the manager should set an incremental reward such that $R_k \sim k^{\alpha}$ with $\alpha > 1.40$, which in turn would worsen the down-payment function both in terms of incremental expenditures and in exploration of bugs with higher ranks. This approach does not consider possible individual human hard limits preventing the finding of additional bugs, which would drive iterative costs $c_k$ sufficiently high as $k$ gets large. In that latter case, setting higher reward incentives would have no effect.\\

Security researchers tend to switch from one bounty program to another program \cite{zhao2014exploratory,zhao2015empirical}. The strategy can be interpreted as portfolio diversification \cite{goetzmann2008equity}. Over all bug bounty programs, security researchers have another much more favorable expected payoff: the reward scaling is smaller ($R_{k} \sim k^{1.24}$ with $ p < 0.001$, $R^2 > 0.99$), yet the frequency of bug discoveries decays much slower as a function of rank $P_{k} \sim k^{-1.05}$ ($ p < 0.001$, $R^2 = 0.85$). Therefore, over all bounty programs, security researchers have an increasing -- yet marginally decreasing -- incentive to explore higher ranks as $U_k \sim k^{0.19}$. In a nutshell, security researchers have an incentive to keep searching for bugs on a large variety of bug bounty programs.

\subsection{Influence of newly launched programs on researcher behaviors}
\label{ols}

As security researchers weigh their strategic choice to switch their attention from one program to another, the time factor is determinant because the expected payoff is dependent on the current vulnerability rank, which maps into the time dimension (i.e, the duration between two discoveries is drawn from a characteristic random variable, which is not considered here). While a researcher may decide to switch at any time, the most obvious moment is when a new program is being launched: incentives shift suddenly and security researchers may decide to leave older programs at the expense of new programs with fresh bug discovery opportunities. However, a number of factors may influence their decision: the reputation of the organization launching the new program (it brings more recognition to submit a bug to e.g., Twitter compared to a less well-known organization), the amount of reward, and the relative time between an old program and the newest one. To encompass the effects of new public bug bounty programs on incumbent programs, we aim to test three hypotheses:
\begin{itemize}
  \item {\bf H1:} An existing bounty program will receive fewer reports when more new programs are launched,
  \item {\bf H2:} An existing bounty program will receive less reports when bounty rewards provided by newly launched programs are higher,
  \item {\bf H3:} The number of newly launched programs has a negative impact on the contribution to older programs.
\end{itemize}

We specify a simple ordinary least square (OLS) regression model as follows:
\begin{equation}
\label{reg_base}
V_{it} = \beta_0 + \beta_1 dP_t + \beta_2 T_{it} + \beta_3 A_{it} + \beta_4 B_{it} + \epsilon_{it}.
\end{equation}
$V_{it}$ is the number of vulnerability reports received by bounty program $i$ in the month $t$. $dP_t$ is the number of new programs launched in month $t$. Hypothesis {\bf H1} predicts that its coefficient ($\beta_1$) is negative. $T_{it}$ is the number of months since bounty program $i$ launched. We consider two control variables that could influence a researcher's decision \cite{zhao2015empirical}. We first incorporate $A_i$ the log of the Alexa rank, which measures web traffic as a proxy of popularity for organization $i$. $B_i$ is the log of the average amount of bounty paid per bug by bounty program $i$. Both $A_i$ and $B_i$ are assumed to remain constant over time. Finally, $\epsilon_{it}$ is the unobservable error term. In models {\bf 2}-{\bf 4}, we extend the basic model (model {\bf 1}) to further study competition occurring between bounty programs. These alternative specifications include:

\begin{itemize}
  \item \textbf{Average bounty of newly launched programs:} intuitively, if new programs offer higher rewards, they should attract more researchers from existing programs. We calculate the average bounty for all new programs in month $t$ as $NB_t$ in models {\bf 2}-{\bf 4}.
  
  \item \textbf{Interaction between $dP_t$ and $T_{it}$:} conceivably, the effect of new programs on existing programs depends on certain characteristics of the latter, such as age. In particular, we ask if a new entrant has more negative effects on older programs compared to younger programs? To examine this, we consider an interaction term between the number of new programs ($dP_t$) and the age of the program ($T_{it}$) in models {\bf 3}-{\bf 4}. Hypothesis H3 predicts that this coefficient should be negative.
  
  \item \textbf{Program fixed effect:} to better control for program-specific, time-invariant characteristics, e.g., the reputation among researchers, we add a program fixed effect in model {\bf 4}. The addition of this fixed effect allows us to examine how bug discovery changes over time within each program $i$.
\end{itemize}

\begin{table}
	\centering
	\caption{Regression results.}
	\begin{tabular}{lcccc} \hline
		& (1) & (2) & (3) & (4) \\
		VARIABLES & $V_{it}$ & $V_{it}$ & $V_{it}$ & $V_{it}$ \\ \hline
		&  &  &  &  \\
		$dP_t$ & -1.235*** & -1.350*** & -2.310*** & -1.236** \\
		& (0.305) & (0.327) & (0.603) & (0.515) \\
		$A_i$ & -23.61*** & -23.72*** & -23.72*** & -7.188** \\
		& (2.140) & (2.156) & (2.152) & (3.473) \\
		$B_i$ & 16.64*** & 16.56*** & 16.75*** & -7.414 \\
		& (1.311) & (1.315) & (1.339) & (5.698) \\
		$T_{it}$ & -0.690 & -0.658 & -3.312*** & -3.758*** \\
		& (0.426) & (0.427) & (1.239) & (1.128) \\
		$B_{new,t}$ &  & -0.0445 & -0.0312 & -0.0321* \\
		&  & (0.0280) & (0.0277) & (0.0184) \\
		$T_{it} \times dP_t$ &  &  & 0.106** & 0.0755* \\
		&  &  & (0.0431) & (0.0406) \\
		Constant & 160.2*** & 170.4*** & 190.3*** & 136.5*** \\
		& (16.12) & (18.80) & (23.17) & (26.17) \\
		&  &  &  &  \\
		Observations & 1,212 & 1,212 & 1,212 & 1,212 \\
		R-squared & 0.314 & 0.316 & 0.319 & 0.647 \\
		Program FE & No & No & No & Yes \\ \hline
		\multicolumn{5}{c}{ Robust standard errors in parentheses} \\
		\multicolumn{5}{c}{ *** p$<$0.01, ** p$<$0.05, * p$<$0.1} \\
	\end{tabular}
	
	\label{tab:reg}
\end{table}

\noindent The regression results are shown in Table~\ref{tab:reg}. Consistent with hypothesis {\bf H1}, the coefficient of $dP_t$ is negative and statistically significant in all four specifications. {\it Ceteris paribus}, the launch of new programs reduces the number of vulnerabilities reported to existing programs. In other words, the entry of new programs indeed attracts researchers' attention away from existing programs, which is consistent with the fast decreasing expected payoff for individuals searching bugs for a specific program. Also, the average bounty paid by new programs ($B_{new,t}$) has a negative effect on existing programs as well, but the coefficient is only significant in model {\bf 4}. Again this result is consistent with the theory and hypothesis {\bf H2}, as researchers have a higher incentive to switch to new programs, if they offer more low-hanging fruits and higher bounties.\\

The interaction coefficients for term $T_{it} \times dP_t$ in models {\bf 3} and {\bf 4} are positive and statistically significant, so they do not support hypothesis {\bf H3}. The result shows that the impact of newly launched programs depends on the age of the existing programs: compared to younger programs, the negative impact of $dP_t$ is smaller for programs with a longer history, i.e., those with larger $T_{it}$. At first sight, this results may look at odds with the fact that individual expected payoff from a specific program decreases as a function of rank $k$, and presumably the older a program the more likely it has a high rank. Thus, the switching effect should be stronger. Perhaps our OLS regression model is limited in the sense that it does account for the absolute activity (which decreases very slowly as $t \rightarrow \infty$, as shown on Figure \ref{timeline}B), instead of the variation rate. Consistent with previous research~\cite{zhao2015empirical}, the regression results also show that a program with higher reputation ($A_i$) or higher bounty ($B_i$) is associated with more bugs received in a month. The regression results also show that a program with higher age ($T_{it}$) is associated with less bugs found. This observation corresponds to the power law decay of bug submission observed following a program launch (c.f., Figure \ref{timeline}B).

% (\textbf{TODO}: We can also test the robustness of the model by trying about alternative dependent variables, such as the the number of researchers that have submitted at least one report to program $i$ in month $t$.)
\section{Discussion}
\label{sec:discussion}
Finding bugs in software code is one of the oldest and toughest problems in software engineering \cite{adams1984textordfeminineoptimizing}. While algorithm-based approaches have been developed \cite{avgerinos2014enhancing}, human verification has remained a prime way for bug hunting. Resorting to the crowd for finding bugs is not new \footnote{The first known bug bounty is the reward check program implemented by Donald Knuth in 1985 for debugging his book {\it TeX: The Program} ({\it http://truetex.com/knuthchk.htm}).}, but bug bounty programs have recently been promoted by the emergence of bug bounty platforms. Here, we have studied the incentive mechanisms across 35 bug bounty programs on HackerOne. Our results show that the number of discovered bugs and vulnerabilities in a bounty program is super-linearly associated with the number of security researchers. However, the distribution of bugs found per researcher per program is bounded: in a given bug bounty program, the marginal probability of finding additional bugs is decreasing rapidly. On the contrary, security researchers have high incentives to switch among multiple bug bounty programs. We find indeed that each newly launched program has a negative effect on submissions to incumbent bug bounty programs. Furthermore, controlling specifically for monetary incentives, we find that the amount of reward for valid bugs in newly launched programs has a negative effect on the number of bug submissions to incumbent programs. These results provide important insights on the theory of bug discovery. They also help draw practical organization design recommendations for bug bounty platforms such as HackerOne, as well as for organizations managing bug bounty programs.\\
 
Our results provide an essential validation step to the theory formulated by Brady et al. \cite{brady1999murphy}. No single security researcher is able to find most bugs in one program. In contrast, a good bug bounty program involves submissions from a diverse crowd of security researchers. Borrowing to the formulation by Brady et al. of software security as a phenomenon of evolutionary pressure dictated by environmental changes, we shall propose that any additional security researcher involved in a bug bounty program brings a unique combination of skills and mindset. This unique perspective is comparable to a slightly changing environment for the software piece under scrutiny, and is associated with unique opportunities for each security researcher, regardless of the opportunity level of other researchers focused on the same software piece. Yet, because people cannot easily change their skills and mindset, once the opportunity has been exploited, finding additional bugs gets much harder. Therefore, researchers tend to turn their attention to newly launched bug bounty programs.\\

The fact that each security researcher has a limited capacity to uncover a large number of bugs for a specific program (i.e., on a specific piece of software) carries a strong justification for the existence of bug bounty programs as a tool for engaging a large and diverse crowd of security researchers, beyond internalized software testing and security research teams. As a concrete case, we discuss the {\it Uber} bug bounty program launched in 2016 \cite{moussouris2016}. Uber has designed its program as a way to select and hire a band of {\it security czars} from a larger crowd. Although there is certainly nothing wrong with hiring security experts in an opportunistic manner, our results rather suggest that systematically using bug bounty programs for hiring may prove counter-productive in the long term. Security researchers will be likely to tune their expectations and behaviors toward getting a job. The approach implemented by Uber may reduce engagement by researchers who do not expect to get a job offer. Thus, it could limit the involvement of a larger crowd and its renewal as more permanent security positions get filled. \\

The strategy followed by Uber for its bug bounty program is also interesting from a theoretical perspective: Uber is following a well-known strategy first described by Ronald Coase \cite{coase1937}, which prescribes that if an organization has repeated interactions on the market with the same counterpart -- or {\it ceteris paribus} a similar counterpart -- then the organization is better off internalizing the resource in order to avoid repeated transaction costs. Hence Uber considers security researchers as {\it substitute goods}, while our results -- and probably the mere existence of bug bounty programs -- rather demonstrate that security researchers are {\it complements}. The distinction between substitutes and complements regarding security researchers brings a fundamental justification for the existence and the future development of bug bounty programs as marketplaces for trading bugs and vulnerabilities \cite{bohme2006comparison}. \\

Yet, the organization designs of bug bounties programs and the online platforms supporting them are still pretty much empirical. The validation step performed here provides additional theoretical insights. On the one hand, these insights shall be useful to formulate recommendations. On the other hand, they help identify blind spots, which will deserve future theoretical and empirical research efforts. We propose three major recommendations that may significantly improve the efficiency of bug bounty programs and the online platforms hosting them, as well as help maximize the engagement of security researchers. 

\subsection{Encourage enrollment, mobility and renewal}
Bug bounty programs shall encourage mobility by devoting resources to the recruitment of security researchers who were not previously involved in the program, rather than increasing efforts to keep security researchers who have already performed well. Mobility increases chances to find security researchers with diverse skills and mindset, who in turn will find additional bugs. Similarly, bug bounty platforms have the possibility to encourage mobility across bug bounty programs. We have found that mobility across programs already exists, in particular mobility from old to newly launched programs. We also advocate the active recruitment of new security researchers by both bug bounty programs and platforms. For example, a bug bounty platform may highlight older programs to security researchers who have recently enrolled on the platform.

\subsection{Feature major changes for front-loading}
The launch of a new bug bounty program is a unique moment, which can attract a large number of security researchers. Yet front-loading may also be organized when a software piece receives a major update with higher probability of finding bugs. For instance, when a new software release contains significant changes in the codebase, bug bounty program managers should feature these changes and help security researchers focus on issues they have not previously been exposed to. This approach may also help bug bounty program managers steer the attention of security researchers toward more pressing security issues. Front-loading can also be organized with a temporary increase of bug bounty rewards. Additionally, dynamic adaptation of incentives can help manage contingencies associated with surges of submissions. For instance, bug bounty managers may decide to reduce rewards during internal overload periods. 
 
\subsection{Organize fluid and low-transaction cost markets}
One overarching concern with trading bugs and vulnerabilities is the temptation by security researchers to consider selling their bugs on the black market. One way to alleviate this problem is to streamline transactions costs associated with bug submission and reward operations. Encouraging mobility without providing market fluidity indeed exposes to the risk that bugs get sold more often on the black market. Our recommendation is reminiscent of the strategy followed by {\it Apple}: by offering cheap enough and easy to download online music, Apple managed to capture most of the online music black market, such as {\it Napster}. Bug bounty programs face similar challenges and opportunities to capture a larger market share by reducing transaction costs and thus offering an alternative to uncertainties associated with the black market.\\

Our recommendations stem directly from the validation step we performed. They show the importance of having a clear view of the theoretical and empirical underpinnings of the mechanisms of bug bounty programs and of platforms organizing them. Mobility, front-loading and market fluidity may be organized either through top-down bureaucracy or by setting market incentives appropriately. The relative advantages of bureaucracy and market organization shall be further studied. Our recommendations for designing bug bounty programs apply indiscriminately to public and private bug bounties. However, we believe that private bug bounty programs face more complex challenges as they select their invited participants. The selection process is costly and {\it de facto} reduces the pool of security researchers.\\

%We also suggests that both individual bug bounty programs and bug bounty platforms shall aim at constantly attracting more researchers, in light of the results in Section~\ref{sec:enrollment}. Previous work~\cite{zhao2015empirical} has made several suggestions, such as a first time bonus, to achieve this goal. We further suggest that current private bounty programs should gradually increase their enrollment and eventually go public. Receiving no bugs in the private stage does not indicate that the system is secure. Rather, it is more likely that the system has not been fully tested by the crowd. In addition, bug bounty platforms should improve their researcher invitation/allocation mechanisms to encourage the flow of diverse researchers to different programs~\cite{zhao2016crowdsourced}.

Our study would benefit from additional empirical research using data which are currently not available from public sources. First, our results are limited by the difficulty to estimate the resource costs that security researchers bear when searching for bugs (e.g., time spent). This information would help further test and understand our results, which show that there are physical limitations regarding the possibility for an individual to find an arbitrarily large number of bugs. Information on the cost functions would also bring further insights on refined expected utility functions by security researchers, in particular, the distinction between expected monetary rewards and effort devoted to reputation seeking.\\

We may also further question how bug bounty program operations impact the motivation of researchers: for example, bug bounty programs may be temporarily overloaded with submissions \cite{zhao2014exploratory,zhao2015empirical}. This overload may stem from priority queueing \cite{maillart2011quantification} and effort required to verify and remediate security incidents internally \cite{kuypers2016empirical}. Delays and contingencies, such as timing and discounting effects, contribute to increase transaction costs and uncertainties for security researchers. Deeper understanding of the dynamics associated with bug bounty program operations may help establish a benchmark on the performance of organization designs and their implementations.\\

Finally, there is evidence that security researchers have specialized knowledge and skills. The competitive environment associated with bug bounty programs reinforces incentives to specialize. At least two types of specialization exist in bug bounty programs: {\it program-specific} and {\it vulnerability-specific} \cite{zhao2014exploratory,zhao2015empirical}. Program-specific specialization is associated with knowledge, experience and skills required to find vulnerabilities in websites and software products in one particular bug bounty program. Since specialization is relatively unique to the program, a specialized researcher has fewer options to switch between programs. Vulnerability-specific specialization is associated with knowledge and skills regarding a particular type of vulnerability, which can exist in many different products. These researchers have stronger incentives to explore different bug bounty programs. Specialization must be accounted for when implementing organization design recommendations. In some circumstances, it may be desirable to attract a crowd of diverse yet specialized security researchers. Depending on the specialization required, targeting specific security researchers may however restrict the diversity of the resource pool. Specialization is directly associated with the concept of {\it skills and mindset}, which we have introduced here to explain the observed hard limits regarding the number of bugs a single researcher can find. This notion deserves a more thorough definition as well as a testable theory.\\

\section{Conclusion}
\label{sec:conclusion}
In this paper, we have investigated how crowds of security researchers hunt software vulnerabilities and how they report their findings to bug bounty programs on dedicated online platforms. Consistent with the famous adage ``Given enough eyeballs, all bugs are shallow'' by Eric Raymond \cite{raymond1999cathedral}, we have found that security researchers face challenging difficulties when trying to uncover large numbers of bugs in the same bounty program: the super-linear reward increase for newly discovered bugs does not counterbalance the sharply decreasing probability of finding new bugs by the same person. This result is consistent with the theory proposed by Brady et al. on maximized entropy of bug discovery as an evolutionary process, following adaptation to changing environments \cite{brady1999murphy}: each security researcher tests software within an environment bounded by her skills and mindset. This result brings a fundamental justification for the existence of markets for bugs, beyond internalized security operations and research: bug bounty programs offer a way to capitalize on these diverse {\it environments} provided by the involvement of many security researchers. Yet, difficulties for researchers to find large numbers of bugs in one bug bounty program bring incentives for mobility across programs. In particular, we find that the launch of new bug bounty programs has a negative effect on incumbent programs regarding bug submissions. We thus propose three organization design recommendations. First, enrollment, mobility and renewal shall be encouraged across bug bounty programs as well as across bug bounty platforms. Second, bug bounty program managers shall highlight major changes in the codebase and get organized for front-loading, in order to help security researchers focus on recent codebase changes. Finally, recognizing that market structures are a powerful mechanism to mobilize large crowds of security researchers, we recommend to organize fluid market transactions, in order to reduce as much as possible uncertainties associated with bug submissions.

%\subsubsection*{Acknowledgements}
\acknowledgements{We thank the anonymous reviewers and editorial staff for their comments, which greatly improved the article. The authors would also like to thank Aron Laszka for his valuable comments. We are further grateful for feedback received at talks at the Stanford Center for Internet Security and Cooperation (CISAC), Telecom Paris Tech, Technical University of Munich, the Higher Moments seminar at the University of Innsbruck, the 2017 CROSSING Conference at the Technical University of Darmstadt, the 2016 Workshop on the Economics of Information Security (WEIS), the 2016 PrivacyCon Conference facilitated by the Federal Trade Commission, the 2016 Legal and Policy Dimensions of Cybersecurity workshop, and the 2017 CRESSE Conference on the Advances in the Analysis of Competition Policy and Regulation. This research was supported in part by the National Science Foundation through award CCF-0424422 (TRUST - Team for Research in Ubiquitous Secure Technology). The research activities of Jens Grossklags are supported by the German Institute for Trust and Safety on the Internet (DIVSI). Thomas Maillart acknowledges support from the Swiss National Science Foundation (SNSF; Grants PA00P2\_145368, P300P2\_158462 and P3P3P2\_167694).}

%}

%\bibliographystyle{apsrev4-1}
\bibliography{references}

\end{document}